# Data-driven Air Quality Characterisation for Urban Environments: a Case Study

**Yuchao Zhou[1], Suparna De[1], Member, IEEE, Gideon Ewa[2], Charith Perera[3], Member, IEEE, and Klaus Moessner[1], Senior Member, IEEE**
[1]Institute for Communication Systems, University of Surrey, UK GU2 7XH
[2]Department of Electronic Engineering, University of Surrey, UK GU2 7XH
[3]School of Computer Science and Informatics, Cardiff University, UK CF24 3AA

Corresponding author: Suparna De (e-mail: s.de@surrey.ac.uk).

This paper describes work undertaken in the context of the TagItSmart! project (www.tagitsmart.eu). TagItSmart! is a collaborative project supported by the European Horizon 2020 programme, contract number: 688061.

**ABSTRACT** The economic and social impact of poor air quality in towns and cities is increasingly being recognised, together with the need for effective ways of creating awareness of real-time air quality levels and their impact on human health. With local authority maintained monitoring stations being geographically sparse and the resultant datasets also featuring missing labels, computational data-driven mechanisms are needed to address the data sparsity challenge. In this paper, we propose a machine learning-based method to accurately predict the Air Quality Index (AQI), using environmental monitoring data together with meteorological measurements. To do so, we develop an air quality estimation framework that implements a neural network that is enhanced with a novel Non-linear Autoregressive neural network with exogenous input (NARX), especially designed for time series prediction. The framework is applied to a case study featuring different monitoring sites in London, with comparisons against other standard machine-learning based predictive algorithms showing the feasibility and robust performance of the proposed method for different kinds of areas within an urban region.

**INDEX TERMS** Air Quality Estimation, Air Pollution, Machine Learning Prediction, Neural Network

## I. INTRODUCTION

With the growing population of the world and the migration of people to urban areas [1], it becomes imperative to create an intelligent and sustainable environment that offers citizens a high quality of life and is geared towards supporting their well-being. The direct effect of this urban drift has had profound effects on social, economic and ecological systems, causing stresses on the environment and society. The social and economic implications include impacts from human activities such as transport, industrialization, combustion, construction etc., all of which have a direct or indirect bearing on the environment. These pollution sources have led to release of pollutants such as Nitrogen dioxide ($NO_2$), Particulate Matter (PM), Sulphur dioxide ($SO_2$) etc. into the atmosphere.

It is recognized that air pollution is influenced by urban dynamics [2]. Recent media reports[1] have highlighted the links between road traffic and large-scale construction activities with toxic air in towns and cities across the UK. Poor air quality has clear public health impacts, with 40,000 deaths annually in the UK (9,500 in London) directly attributable to air pollution and exacerbating health conditions with those with heart or lung conditions [3]. Spikes in air pollution levels have also been directly linked with increased hospital and GP visits [4], pointing to additional costs faced by the public health service in treating conditions exacerbated by poor air quality. This calls for effective ways of creating awareness of real-time air quality levels and their impact on human health.

Since air pollution is highly location dependent [2] and air quality monitoring stations installed at fixed-site stations, though very accurate, have high installation costs, are bulky and geographically sparse (the UK's DEFRA Automatic Urban and Rural Monitoring Network (AURN) has 168 sites covering the entire UK [5]), this poses challenges for evidence-based real-time air quality-related decision making, both for city authorities and citizens. Secondly, the data from these monitoring stations has lots of missing labels due to the maintenance schedules of the devices in the station [6].

---
[1] https://www.theguardian.com/environment/2018/aug/28/too-dirty-to-breathe-can-london-clean-up-its-toxic-air

Since there are a large number of air pollutants, which can combine actively or reactively to form secondary pollutants, countries have adopted the Air Quality Index (AQI) as a measure of pollutants in the air. It is an easily understandable value that shows how polluted the air is or how polluted it will be in future. This information can be used to warn the public or sensitive groups about the state of pollution of the environment.

Beginning with the first use in Toronto in 1969, AQI calculation and prediction has gained popularity and is widely adopted by many countries [7]. The complexity and number of factors affecting the AQI has motivated the use of computational intelligence techniques in the prediction of air quality, achieving higher accuracy than statistical methods such as moving average or linear or Gaussian interpolation [8]. The emerging paradigm of urban computing [9], which aims to analyse the correlations and patterns from urban big data to infer unknown knowledge [10], has researched various aspects of air pollution, for instance, by employing data-informed air quality prediction algorithms (to mitigate the data sparsity challenge [11]), with the developed Machine-Learning (ML)-based algorithms achieving a high performance in terms of the prediction accuracy and efficiency [8, 12, 13]. Most of these research works implement techniques to predict and identify patterns relevant to individual pollutant concentrations, for example, $PM_{2.5}$ [6, 14], Carbon Monoxide (CO) [12, 13, 15], $PM_{10}$ [8, 16], Nitrogen Oxides ($NO_x$) [8, 15-17]. Other allied works seek to employ supervised methods that take into account historical AQI values in order to perform short-term predictions of AQI measures for the same or neighbouring regions [18, 19].

However, it has been noted that there should be three stages involved in predicting AQI [20]: 1) establishment of Air quality model, 2) identification of meteorology factors and forecast, and 3) doing the actual AQI forecast and estimation based on identified algorithms. The AQI calculation model choice is important since pollutants vary from place to place, for example, an urban area may be concerned about $NO_2$ because of large vehicular presence, an industrialized area might want to monitor $SO_2$ and a city like Madrid may be interested in pollen because of its prevalence in this region. Thus, the AQI model needs to consider individual pollutants or a combination of them. Meteorology is an influencing factor since it has been established that factors such as temperature, atmospheric pressure, relative humidity, wind speed and wind direction are dominant factors that influence pollutant concentration and by extension AQI [16].

To implement the requisite three phases and to address the data sparsity and unlabeled data challenges, this paper sets out a comprehensive air quality estimation framework that implements an AQI model encompassing a predictive algorithm for air quality index, given pollutant and meteorology data. The novel predictive method applies the Non-linear Autoregressive neural network with exogenous input (NARX) time series prediction model that considers meteorological inputs and previous pollutant values. The selected AQI calculation model also proposes and evaluates two approaches for AQI characterization and prediction: the first of which trains the NARX algorithm directly on the calculated historical AQI values, and the second predicts individual pollutant values before feeding them into the AQI calculation model. Evaluations based on a real-world dataset, and comparison to the state-of-the-art methods in terms of standard evaluation metrics, i.e., Root Mean Squared Error, Mean Absolute Percentage Error, and Band Accuracy, show the feasibility and performance improvements achieved from the proposed approach.

The rest of the paper is organised as follows. Section 2 provides a review of the related work and techniques for AQI and pollutant estimation. The details of the AQI calculation model and meteorology factors characteristics are described in Section 3. Section 4 presents the AQI estimation framework, including algorithmic details of the NARX predictive model. Section 5 presents the experiments performed on a dataset collected from a real-world deployment of monitoring sites across several boroughs of the city of London and also discusses the evaluation results based on the standard metrics by comparing to existing methods. Section 6 concludes the paper and outlines the future research directions.

## II. RELATED WORK
Prediction of air quality levels is important for communicating pollution risks and exposure level. However, it is a complex measure to calculate since the form and dispersal patterns of pollutants are affected by environmental and meteorological factors. The early approach was human-centred, where data collected from different monitoring stations were evaluated based on human experience; hence, making it unreliable. Currently, computational intelligence approaches involve use of smart algorithms such as decision trees, neural networks, self-organizing maps, support vector machines etc. in predicting air quality. This method is advantageous because of its high accuracy and computational efficiency [21].

Zhang et al [22] identified the major techniques for AQI forecasting to include simple empirical approach, and statistical approach. The empirical approach is based on persistence, which factors in current AQI into the prediction of future AQI since it assumes that the current pollutant value has a direct effect on tomorrow's predicted value. This approach is simple and good for stationary conditions but can't handle sudden changes in pollutant and weather. Statistical approach relies on the fact that weather and pollutant concentrations are related statistically i.e. there is correlation between these two elements and therefore regression and trained neural network functions are employed to forecast pollutant concentration. Zhang et al. [22] mention the common algorithms to include Classification And Regression Tree (CART), Artificial Neural Network (ANN), and fuzzy logic. Their work noted that ANN has fast computational speed and an ability to learn and adapt itself to new instances.

**Machine learning-based approaches:** Moustris et al. [15] applied an ANN model for short-term forecasting of $SO_2$, $NO_2$, Ozone ($O_3$) and CO levels across seven monitoring sites in Athens, with evaluation statistics showing a good agreement between predicted and observed pollutant values. The study concluded that ANN can be used effectively for time series prediction and is optimized for problems with big state variables or large dimensions. Hourly concentration of $NO_2$ and NO and meteorology were used in [17] to forecast their values using neural network and Support Vector Machine (SVM), with SVM's ability to set the size of the hidden layers automatically providing better performance than ANN. Another finding from this was that factor-less prediction i.e. prediction without external variables, is fine but additional external variables greatly improve prediction. The downside of this is that if the external variables are predicted, then it could worsen the performance of the algorithm due to accumulated prediction error. The use of ANN for hourly prediction of pollutants was also demonstrated in [16], with known pollutant concentration values at 1, 2 and 3 hour, respectively, prior to the prediction, used to approximate the impact of background factors such as industrial, restaurant and resident emissions. This method was used to predict pollutant concentrations an hour in advance. Comparison of this ANN-based method with multiple linear regression models shows that regression models perform better for predicting CO and $PM_{10}$ values, with mixed results for $NO_2$ (comparable performance) and $O_3$ (ANN performs markedly better). The authors also introduced an 'unknown-background' ANN method, where the predicted concentrations were used as background factors for the following hour prediction, resulting in improved performance for the ANN method. Grid-based forecasting of $PM_{10}$ levels using ANN for a spatial classifier that co-trains a semi-supervised model with spatial features such as points-of-interest density and highway length, was used in [8]. This was extended with a temporal classifier based on conditional random field that considered temporal features such as traffic and meteorology. To address the problem of data sparsity from geographically sparse air quality monitoring stations installed by government agencies, HazeEst [13] and the work in [12] combined the data from static sites with mobile sensor data to forecast CO values for the metropolitan area of Sydney by training and evaluating a number of regression models. Their findings show that SVR has the same estimation accuracy as decision tree regression, but higher than multi-layer perceptron and linear regression.

**Deep Learning approaches:** Recent studies [6, 14] have investigated the use of different deep learning neural networks to perform forecasting of pollutant concentrations. The Deep Air Learning (DAL) model [6] uses a sparse auto-encoder to impose sparsity constraints on the input units to enable the irrelevant input features to be ignored and the main features relevant to the target to be explicitly revealed for association analysis. The deep neural network-based approach in [14] uses a spatial transformation component for spatial correlation and a distributed fusion network to merge all the influential factors for $PM_{2.5}$ forecasting.

**Urban Computing approaches:** Allied research on transport-related themes has considered the impact of weather changes on predicting traffic levels at different points in a city [23], and predicting transport carbon emissions within a city [24]. Recent studies have explored urban models to predict air quality in city districts by considering a range of spatio-temporal urban big data sources such as meteorology, vehicular traffic and points of interest (POI) [2]. It is worth noting that different cities and their public spaces are characterised differently based on their specific natural and built environment [23], which needs to be considered while calculating and predicting the pollution index and discovering the latent temporal and spatial patterns.

From the review of existing works, it is apparent that several authors have used neural networks in their work to model and predict air quality and pollutant concentration. The choice of this machine learning algorithm is strongly based on its fast-computational attributes and its ability to learn and adapt to new instances. Hassan et al. [25] noted that air quality prediction has complex and non-linear patterns. These patterns of data can be efficiently handled by neural networks. Additional features in air quality prediction increase the dimension of data, and Hassan et al. stated that ANN is naturally suited for problems with large number of state variables. Neural networks' ability to make generalizations given an input and its non-mapping capability makes it a good tool for time series prediction. Thus, in this work, we explore a neural network-based algorithm and incorporate a time delay to take into account prior pollutant concentrations into the prediction of future AQIs. Compared to the existing works, our work considers individual pollutant concentrations to provide a comprehensive AQI characterisation and prediction framework.

## III. BACKGROUND

In this section, we first establish the adopted AQI calculation model, setting out how to calculate Air Quality Index (AQI) based on the collected dataset. The characteristics of the sensing sites that are used as the data sources are then presented and analysed. Then we present the statistics of the collected meteorological and pollution data.

TABLE I
INFORMATION OF SENSING SITES

| Borough | Site | Site Type |
|---|---|---|
| Barking and Dagenham | Rush Green | Suburban |
| Bexely | Belvedere West | Urban Background |
| Bexely | Erith | Industrial |
| Reigate and Banstead | Horley | Suburban |
| Reigate and Banstead | Poles Lane | Rural |
| Richmond Upon Thames | Ntl Physical Lab | Suburban |
| Westminster | Marylebone Road | Kerbside |

## A. AQI CALCULATION

This section sets out the adopted AQI calculation model, which is the first stage for AQI estimation for an urban region.

Choosing an appropriate model for representing AQI is challenging. A common and widely used model is that by the United States (US) Environmental Protection Agency (EPA), which identifies six major pollutants as AQI indicators. These include $NO_2$, $CO$, $O_3$, $SO_2$, $PM_{2.5}$ and $PM_{10}$. The EPA model has widely been adopted by many countries, with slight modifications on the pollutant threshold level. The Department of Environmental and Food Research Agency (DEFRA) model is only applicable in the United Kingdom as it does not factor in CO in the AQI calculation. This is because of the steady decrease in carbon monoxide emissions in the UK over the past decade, due to decrease in CO emission sources such as road transport, iron and steel production and in the domestic sector as well [26]. On the other hand, the Common Air Quality Index (CAQI) proposed for use in Europe, which uses the same interpolation formula as the EPA model for calculating the individual AQI of pollutants, has a low tolerance of pollutants. This limits its applicability to serve as the basis of a warning system in countries outside Europe.

In this paper, we adopt the EPA model for AQI calculation. This is because it can be applied across diverse regions, with a single pollutant concentration or a combination of two or more of these enough to compute AQI. As a result, the model enables the pollutants of interest in an area to be considered and also allows for different pollutants to form the key determinant for the AQI of that region, which may be the case due to the specific natural and built environment of that region.

To compute AQI using the EPA model, the concentration of pollutants is measured and their Individual Air Quality Index (IAQI) is computed using the formula in equation 1, as given in [27]. The highest IAQI value becomes the AQI and the pollutant with the highest AQI becomes the key pollutant:

$$AQI_p = \frac{I_{Hi} - I_{Lo}}{BP_{Hi} - BP_{Lo}} \times (C_P - BP_{Lo}) + I_{Lo} \qquad (1)$$

TABLE II
DATA STATISTICS OF SENSING SITES

| Site | Meteorological Data | | | | | | | Pollutant Data | | | | | | AQI Dominant | Dominating Rate (%) | Missing Rate (%) |
|---|---|---|---|---|---|---|---|---|---|---|---|---|---|---|---|---|
| | Temperature | Wind Speed | Wind Direction | Rainfall | Humidity | Solar Radiation | Pressure | $NO_2$ | $PM_{10}$ | $O_3$ | $PM_{2.5}$ | $CO$ | $SO_2$ | | | |
| Rush Green | Y | Y | Y | Y | | Y | Y | Y | | | | | | $NO_2$ | 100 | 14 |
| Belvedere West | Y | Y | Y | | Y | | | Y | Y | Y | Y | | | $O_3$ | 69 | 20 |
| Erith | Y | Y | Y | | | | | Y | Y | | Y | | | $PM_{10}$ | 90 | 50 |
| Horley | Y | Y | Y | Y | Y | | | Y | Y | | | | | $PM_{10}$ | 80 | 37 |
| Poles Lane | Y | Y | Y | | | | | Y | | Y | | | | $O_3$ | 86 | 6 |
| Ntl Physical Lab | Y | Y | Y | | | | | Y | | Y | | | | $O_3$ | 88 | 41 |
| Marylebone Road | Y | Y | Y | | | | | Y | Y | Y | | Y | Y | $NO_2$ | 71 | 34 |

Where $AQI_p$ is the index for pollutant p, $C_P$ is the truncated concentration of pollutant p, $BP_{Hi}$ is the concentration breakpoint that is greater than or equal to $C_P$, $BP_{Lo}$ is the concentration breakpoint that is less than or equal to $C_P$, $I_{Hi}$ and $I_{Lo}$ are the AQI values corresponding to $BP_{Hi}$ and $BP_{Lo}$ respectively.

This model further converts the pollutant concentrations to a number on a scale of 0 to 500. Any number in excess of 100 is considered unhealthy. This is further subdivided into six categories namely "0-50", "51-100", "101-200", "201-300", "301-400", "401-500", with different countries having slight differences in the breakpoints for the above categories, which denote different levels of health concerns, ranging from Good (0-50) to Hazardous (>301).

## B. AIR QUALITY MONITORING SITE CHARACTERISTICS

LondonAir[2], the London Air Quality Network (LAQN) website, provides the datasets from the large-scale deployment of air pollution monitoring sites across London. Sensing sites are deployed on different kinds of areas, with the designated types covering: Urban Background, Industrial, Rural, Suburban, and Kerbside. As different kinds of sites measure different observations, the sites in Table I are selected as both pollution and meteorological data are monitored and accessible from these sites. These seven selected monitoring sites are located in five boroughs of London. The framework developed in this paper has been applied to real data sources obtained in London, UK, and contains the following datasets: meteorological: temperature, wind speed, wind direction, rainfall, humidity, solar radiation and barometric pressure, collected every hour; air pollutants: real valued concentrations of six kinds of pollutants, consisting of $NO_2$, $PM_{10}$, $PM_{2.5}$, $CO$, $SO_2$ and $O_3$, reported by the ground-based monitoring stations every hour. The datasets were collected over a number of years (2013-17), covering the first five months of the year, i.e. January to end of May (inclusive), since we found these months to have the most complete datasets.

---
[2] https://www.londonair.org.uk/LondonAir/Default.aspx

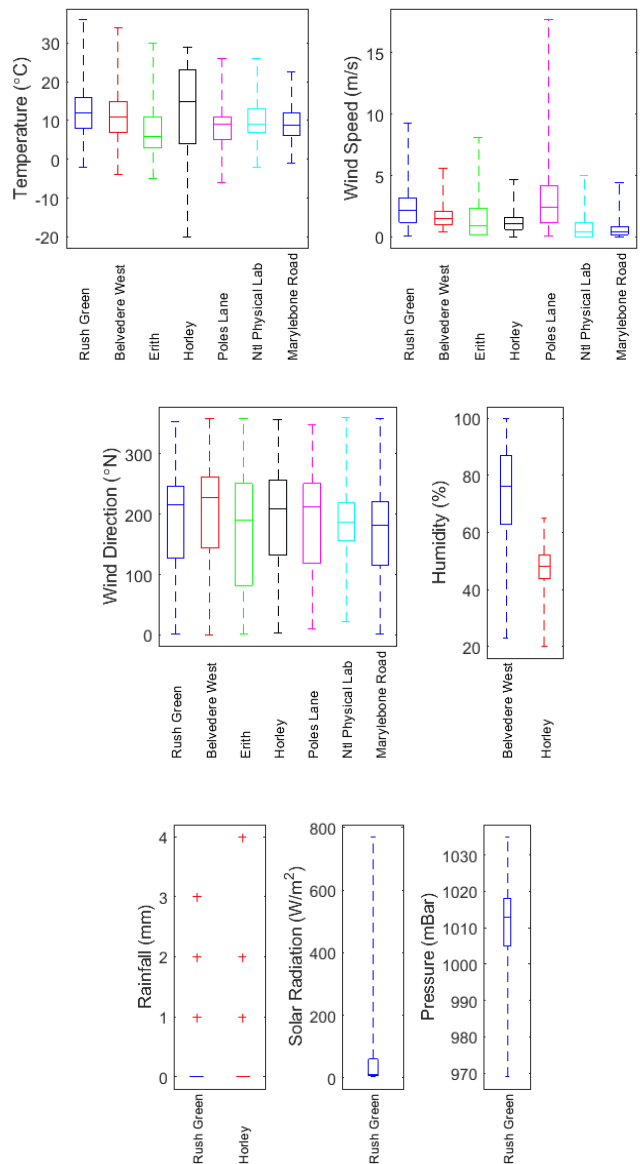

**FIGURE 1.** Boxplot Comparing the Distribution of Different Meteorological Features for the London Monitoring Stations.

As shown in Table II, all the monitoring sites report data for temperature, wind speed, wind direction, and $NO_2$. The other observations are measured by some of the sites. The dominant pollutants are $NO_2$, $O_3$, and $PM_{10}$ across the different sites. The dominant rate is derived by calculating the percentage of how many times the pollutant dominates in the calculation of the AQI of the area over the total number of measured records. It is apparent from the statistics in Table II that the datasets have missing records, for simplification, these rows are removed during the data cleaning stage of the experiments. However, this approach may result in some meaningful data being omitted. To overcome this problem, missing data estimation approaches, as evidenced in our previous work [11], can be applied at pre-processing step to obtain a complete dataset. Our approach simply assumes this step has already done and the training dataset is ready to be processed by the approach.

### C. POLLUTANTS AND METEOROLOGY

Figure 1 shows the boxplots of the meteorological data of the different sensing sites. Except for the monitoring site of Horley, the temperature data shows a similar pattern for the different areas even in different years. This shows that there are small variations in temperature values in the inner boroughs of London, where the monitoring sites are located, over the winter and spring seasons for the evaluated years. The temperature data for Horley shows a median higher than that recorded at the other sites, but also containing extremely low minimum temperature values of -20 °C, which might be attributed to the data containing outliers. Wind speed does not vary too much, with the median range from 1 to 2 m/s. However, the Poles Lane monitoring site reported some wind speed measurements much higher than that from the other sites. A possible reason for this is that the site is a rural area and may not have a substantial built environment near the site, which can act as an obstacle to the wind. Wind direction shows stable distributions across all sites. Wind direction was measured within a 360° angle (i.e. all directions) and the measurements were mostly dominated by one direction, i.e. around 200° to the north. Rainfall is reported by only two of the selected sites in the datasets. Most of the data is composed of 0 values and several of them are 1, 2, 3, and 4 mm. Humidity is also measured by two sites; however, there is a large difference in the measured values, with the 'urban background' site of Belvedere West reporting higher humidity values than that of the suburban site in Horley. Solar radiation and pressure are only available for the Rush Green site; thus, it cannot be compared to the others.

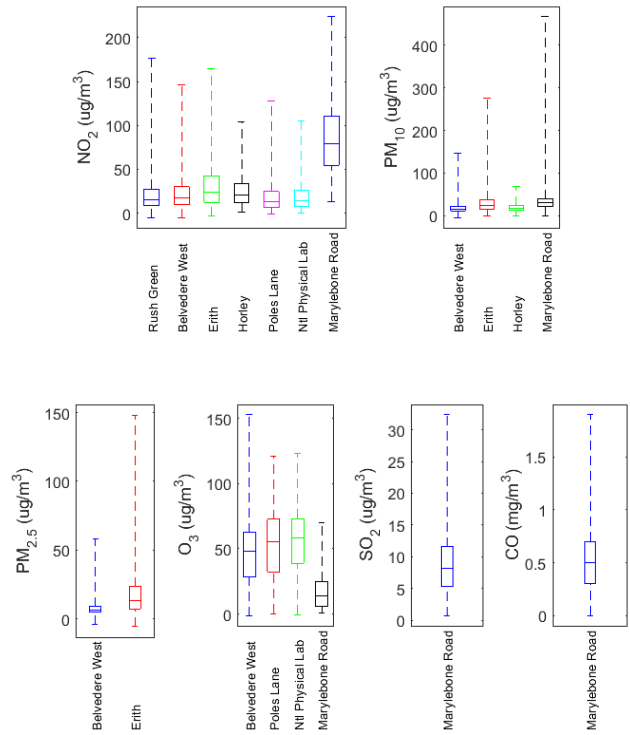

**FIGURE 2.** Boxplot Showing the Distribution of Individual Pollutant Concentrations for the Different London Monitoring Stations.

Figure 2 provides the boxplots of the measured pollutants values. $NO_2$ is reported from all of the selected sites. $NO_2$ values at the kerbside site of Marylebone Road are much larger than those from the other sites. This is because $NO_2$ is mostly generated by road traffic and corresponds to the kerbside location of this sensing site and the urban nature of this location. On the contrary, Marylebone Road has lower $O_3$ values than those reported at the other sites, pointing to a possible inverse correlation; because $O_3$ is a secondary pollutant formed by the reaction of $NO_x$ with hydrocarbon under ultraviolet light. The other observations of $PM_{10}$ and $PM_{2.5}$ show similar distributions but differences in the extreme values. For example, Marylebone Road contains high $PM_{10}$ values, while Erith has large values reported for $PM_{10}$ and $PM_{2.5}$, pointing to a link to its industrial location. $CO$ and $SO_2$ are only measured at the Marylebone Road site in our datasets. These two pollutants show low concentrations at this site and are not considered the main source of pollution in London.

Figure 3 shows the AQI distributions of the different sensing sites. Calculated AQI values of Rush Green and Horley show low values throughout, with more than 75% falling within the 'Good' band and the maximum AQI value in the Moderate band. The AQIs of Belvedere West, Erith, Poles Lane, and Ntl Physical Lab show a larger variance than the previous two sites. Although most of them are within the ranges of the Moderate and Good bands, some values are high and extend to the 'Unhealthy' and 'Very Unhealthy' bands. For the kerbside Marylebone Road site, most values are Good or Moderate, but the maximum calculated AQI reaches the 'Hazardous' range.

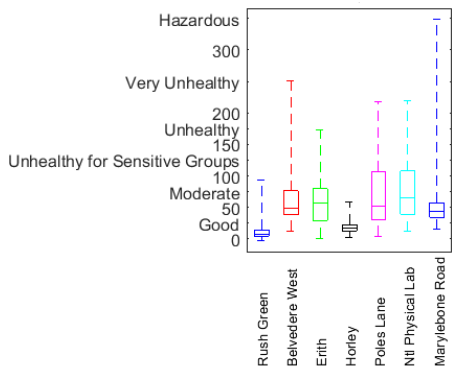

**FIGURE 3.** Boxplot Comparing the Air Quality Index Distributions for the Different London Monitoring Stations.

## IV. AIR QUALITY ESTIMATION FRAMEWORK

Figure 4 presents the proposed air quality estimation framework, which combines meteorological data as well as pollutant data with a one-step temporal delay to provide estimates of AQI values. The two approaches developed in this work are shown in Figure 4. Both approaches begin with a data cleaning phase. The left-hand side of Fig. 4, which depicts the first approach developed in this work for AQI estimation, *AQIPredict*, computes AQIs based on the original pollutant concentrations. It then trains a prediction model that applies meteorological data and the previously calculated AQIs to predict AQIs. On the other hand, the right-hand side of Fig. 4, which shows the second approach being proposed in this work, *Pollutant2AQI*, trains a prediction model directly with the meteorological data and the previous pollutant values to predict pollutant values. The individually predicted pollutant values are then used to compute the final estimates of AQI values.

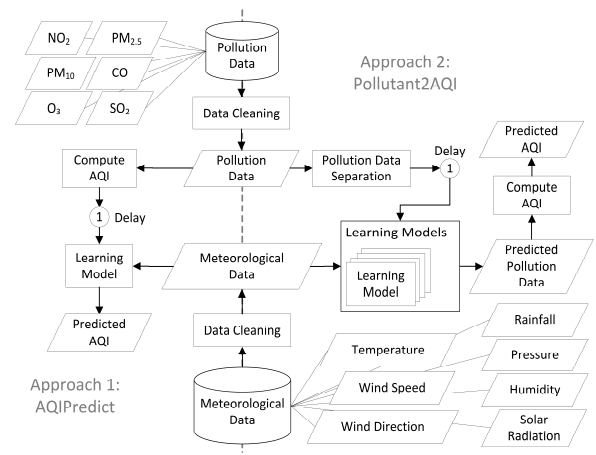

**FIGURE 4.** Air Quality Estimation Framework.

The Learning Model in the framework applies a Nonlinear Autoregressive Neural network with eXogenous input (NARX) [28, 29] to provide time series pollution data/AQI prediction with meteorological data as exogenous input. NARX is based on recurrent dynamic neural network, which has a memory of its previous state. The NARX will learn a function of equation:

$$y(t) = f(y_{t-d}, \mathbf{x}_{\mathrm{meteorological}}) \qquad (2)$$

where $y_{t-d}$ is the previous value of y and d is the output time delay (1 in our experiments), $\mathbf{x}_{\mathrm{meteorological}}$ is a vector of meteorological data.

The NARX can be trained by steepest descent algorithm, Newton's method as well as Levenberg Marquardt (LM) algorithm [30, 31]. LM algorithm is applied in our framework and introduced below. The aim of the training is to get the weights for least square error. The sum of squared error of NARX is defined as a function $E(\omega)$ of weights vector $\omega$ with $N$ samples.

$$E(\omega) = \frac{1}{2}\sum_{q=1}^{N}(e(\omega))^2 \quad (3)$$

The Gauss-Newton method provides a solution of changing weights $\Delta\omega$ for a step as follows:

$$\Delta\omega = -\left[\nabla^2 E(\omega)\right]^{-1}\nabla E(\omega) \quad (4)$$

where $\nabla^2 E(\omega)$ is the Hessian matrix and $\nabla E(\omega)$ is the gradient, which can be calculated by following equations:

$$\nabla^2 E(\omega) = J^T(\omega)J(\omega) + S(\omega) \quad (5)$$

$$\nabla E(\omega) = J^T(\omega)e(\omega) \quad (6)$$

where $J(\omega)$ is the Jacobian matrix of size $N \times P$, $P$ being the size of $\omega$;

$$J(\omega) = \begin{bmatrix} \frac{\partial e_1(\omega)}{\partial \omega_1} & \cdots & \frac{\partial e_1(\omega)}{\partial \omega_p} & \cdots & \frac{\partial e_1(\omega)}{\partial \omega_P} \\ \vdots & \ddots & \vdots & \ddots & \vdots \\ \frac{\partial e_q(\omega)}{\partial \omega_1} & \cdots & \frac{\partial e_q(\omega)}{\partial \omega_p} & \cdots & \frac{\partial e_q(\omega)}{\partial \omega_P} \\ \vdots & \ddots & \vdots & \ddots & \vdots \\ \frac{\partial e_N(\omega)}{\partial \omega_1} & \cdots & \frac{\partial e_N(\omega)}{\partial \omega_p} & \cdots & \frac{\partial e_N(\omega)}{\partial \omega_P} \end{bmatrix} \quad (7)$$

and

$$S(\omega) = \sum_{q=1}^{N} e_q(\omega)\nabla^2 e_q(\omega) \quad (8)$$

Gauss-Newton method assumes $S(\omega) \approx 0$, thus,

$$\Delta\omega = \left[J^T(\omega)J(\omega)\right]^{-1}J^T(\omega)e(\omega) \quad (9)$$

while the LM algorithm makes the following modification to it:

$$\Delta\omega = \left[J^T(\omega)J(\omega) + \mu I\right]^{-1}J^T(\omega)e(\omega) \quad (10)$$

where $I$ is an identity unit matrix and $\mu$ is a parameter controlling the size of the trust region. When $\mu$ is large, the method turns into a steepest descent method with a small step size $1/\mu$, whereas it turns into Gauss-Newton method when $\mu = 0$. If one step reduces overall error, $\mu$ is divided by a factor $\beta$. Otherwise, $\mu$ is multiplied by the

Algorithm 1. LM Training

1. **INPUT**: Training dataset $d$
2. **OUTPUT**: Converged network *net*
3. Compute outputs of the network *net* based on the inputs in $d$ using Equations (12) and (13)
4. Compute the sum of squared errors $E$ of *net* using Equation (3)
5. Compute the Jacobian matrix $J$ using Equations (15) (14) (11) and (7)
6. Get changing of weights $\Delta\omega$ using Equation (10)
7. Compute sum of squared errors $E_{new}$ of a network using new weights $\omega_{new} = \omega + \Delta\omega$
8. **IF** $E_{new} < E$
9.    Reduce $\mu$ in Equation (10) by $\beta$
10.    Apply $\omega_{new}$ to *net*
11.    **IF** converged
12.      Stop and return *net*
13.    **ELSE**
14.      Repeat from Line 3
15.    **END IF**
16. **ELSE**
17.    Increase $\mu$ by $\beta$,
18.    Repeat from Line 6
19. **END IF**
20. The algorithm is converged when the norm of the gradient $\nabla E(\omega)$ (Equation (6)) is less than a predefined value, or when the sum of squared errors $E$ has been reduced to a certain error goal.

factor. By defining $\delta_i^k = \frac{\partial e_q(\omega)}{\partial net_i^k} = f'(net_i^k)$, the elements in Jacobian matrix can be written as

$$J_{q,p} = \frac{\partial e_q(\omega)}{\partial \omega_p} = \frac{\partial e_q(\omega)}{\partial \omega_{i,j}^k} = \frac{\partial e_q(\omega)}{\partial net_i^k}\frac{\partial net_i^k}{\partial \omega_{i,j}^k} = \delta_i^k o_j \quad (11)$$

where q is the q$^{th}$ sample, p is the p$^{th}$ weight, $\omega_{i,j}^k$ indicates the weight connects unit j to unit i in the k$^{th}$ layer, $net_i^k$ is the input of unit i in the k$^{th}$ layer, and $o_j$ is the output of unit i from unit j in the (k-1)$^{th}$ layer. The relations of them are:

$$net_i^k = \sum_{j}^{M_{k-1}} \omega_{i,j}^k o_j^{k-1} + b_i^k \quad (12)$$

where $M_{k-1}$ is the number of units in layer k-1; and

$$o_i^k = f(net_i^k) \quad (13)$$

This can be computed by backpropagation algorithm

$$\boldsymbol{\delta}^{\mathbf{k}} = f'(\mathbf{net}^{\mathbf{k}})\boldsymbol{\omega}^{\mathbf{k+1}^T}\boldsymbol{\delta}^{\mathbf{k+1}} \quad (14)$$

where $f'(\mathbf{net}^k)$ is the derivative of function in of a unit in layer k with respect to its input, with a modification at the final layer.

$$\delta^L = -f'(\mathbf{net}^L) \qquad (15)$$

where *L* indicates the final layer.

Algorithm 1. LM Training describes the process of training a neural network with LM algorithms. Given a Training dataset *d*, LM algorithm iteratively adapts weights in the network until it is converged. In the first iteration, it calculates outputs of an initial network net based on Equations (12), (13), and inputs in *d* (Line 3). With those outputs and original outputs in *d*, the sum of squared errors *E* can be obtained according to Equation (3) (Line 4). The algorithm then computes the Jacobian matrix and gets changing of weights of *net* (Line 5-6). New weights are calculated and applied to a network to compute sum of squatted errors $E_{new}$ based on *d* (Line 7). If $E_{new}$ < *E*, in Equation (10) is reduced by , the new weights are applied to the *net* to continue the next iteration (from Line 3); otherwise in Equation (10) is increased by , the algorithm re-computes (from Line 6) changing of weights of *net* and compares new errors with *E* (Line 8-19). During this checking, if the algorithm converged under the condition at Line 20, the final trained *net* is returned.

## V. EXPERIMENTS AND RESULTS

To evaluate our proposed AQI estimation methods, we design experiments to compare the two proposed approaches for AQI prediction introduced in Figure 4 with different learning algorithms, i.e., Linear Regression (LR) [32], Logistic Regression (LoR) [33], SSVR [34, 35], and NARX [30, 31], with the datasets described in Section III. The algorithms are implemented using the Statistics and Machine Learning Toolbox and Deep Learning Toolbox in Matlab R2017b. The NARX neural network applies 10 hidden layers. The meteorological data are set without any time delay while the pollution data/AQIs are set with one-step time delay. The experiments randomly choose 75% data for training and 15% for testing. For the proposed NARX-based method, another 15% are used for validation. All the methods are performed 10 times and evaluated by using the mean values of the following evaluation metrics: Root Mean Squared Error (RMSE), Mean Absolute Percentage Error (MAPE), and band accuracy. RMSE and MAPE are calculated as per equations 15 and 16, and band accuracy is the percentage of how many predicted AQIs are in the same band of actual AQIs over the total number of data points in the test set.

$$RMSE = \sqrt{\frac{1}{n}\sum_{i=1}^{n}(\hat{y}_i - y_i)^2} \qquad (16)$$

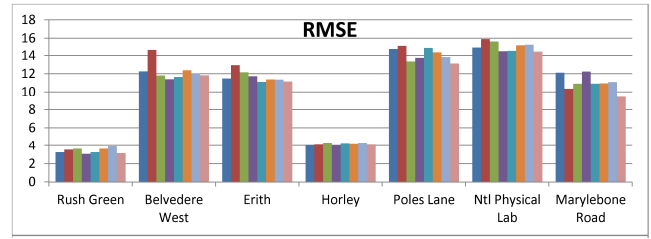

(a) Root Mean Squared Error (RMSE)

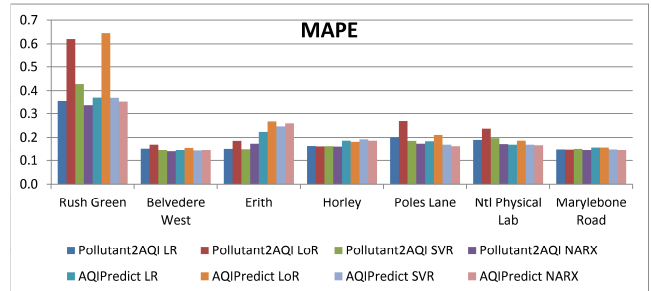

(b) Mean Absolute Percentage Error (MAPE)

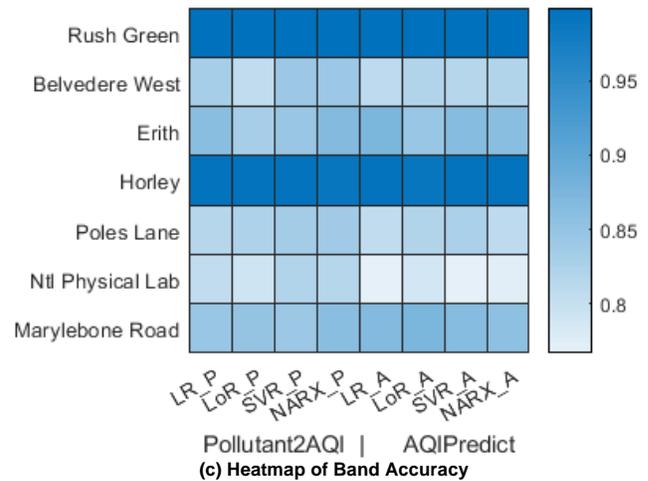

(c) Heatmap of Band Accuracy

**FIGURE 5.** Results of AQI Prediction of Different ML Approaches.

$$MAPE = \frac{100}{n}\sum_{i=1}^{n}\left|\frac{y_i - \hat{y}_i}{y_i}\right| \qquad (17)$$

where *n* is the number of data points in the test set; $\hat{y}_i$ is the predicted value for the *i*th input, and $y_i$ is the corresponding target value.

### A. AQI PREDICTION: RESULTS AND DISCUSSION

In the results' diagrams, we use *AQIPredict* to indicate Approach 1 that uses meteorological data and historical values of AQI (calculated from the individual pollutants' concentrations using Eq. 1, prior to training) to predict future AQI values. We use *Pollutant2AQI* to present Approach 2 that uses meteorological data and the historical pollutants values to predict individual pollutant values and then computes the AQIs based on predicted values, using Eq. 1.

Figure 5 (a), (b) and (c) show the results for RMSE, MAPE and band accuracy, respectively, for the predicted AQI values. It is clear that the results vary a lot across the different sensing sites. This is because firstly, the different monitoring sites are sited differently (e.g. kerbside vs. rural location) and located in different kinds of areas which have different meteorological and pollution characteristics. Secondly, these sensing sites measure different meteorological and pollution data, thus features of the model are different between different sites. Thirdly, pollutants' concentrations are dispersed differently and dominate different areas, depending upon on a number of factors such as industrial activities, vehicular emissions, human activities such as construction, etc.

According to Table II, Rush Green is a site recording six kinds of meteorological data but only one type of pollution data: $NO_2$. Its AQI in Fig. 3 shows that the pollution values range from 0 to around 100 and most of them are below 25, i.e., the AQIs are always in the 'Good' band. For these reasons, all the methods perform well on this dataset achieving a band accuracy of close to 100% (over 99.6%, see Fig. 5c). With respect to RMSE and MAPE, the proposed NARX methods perform the best on both approaches. It is worth noting that even though the RMSE values do not show much difference between the evaluated machine learning algorithms, the MAPE values of LoR on both *AQIPredict* and *Pollutant2AQI* are much worse than the others. This is due to the fact that the AQI data values from Rush Green are small, hence, a small number of errors may not reflect much on the RMSE value but may show up in the MAPE which is significantly affected when the calculation involves the ratio of actual small values.

Another similar sensing site is Horley, which records four meteorological features and two pollutants' data: $NO_2$ and $PM_{10}$ (with $PM_{10}$ the dominant pollutant). The mean values of AQIs of this site are slightly higher than that of Rush Green, nevertheless, almost all the AQIs fall within the 'Good' band. Hence, the band accuracies of predicted values from this site are also close to 100 percent (over 99.1%, see Fig. 5c). RMSE and MAPE values are low for all the methods. RMSE values are close to each other as shown in Fig. 5a, but the MAPE results of the *Pollutant2AQI* methods are less than those of *AQIPredict* methods. Among them, the proposed *Pollutant2AQI* NARX method performs the best for both evaluations. For band accuracy, *Pollutant2AQI* NARX reaches an accuracy of 99.13%, slightly less than the best achieved result of 99.42% obtained by *Pollutant2AQI* LR and *Pollutant2AQI* LoR.

Belvedere West is a site with four meteorological features and four kinds of pollution data: $NO_2$, $PM_{10}$, $O_3$ (dominant pollutant), and $PM_{2.5}$. AQIs of this site ranges from 0 to around 250, covering five bands. Most of the AQIs are located in the Good and Moderate bands. With regards to the evaluation results for this site, *Pollutant2AQI* NARX performs the best for all three metrics.

The Erith sensing site monitors three meteorological features and three kinds of pollutants: $NO_2$, $PM_{10}$ (dominant), and $PM_{2.5}$. The AQIs of this site range from 0 to around 170, covering four bands, with the majority of the AQI values falling within the Good and Moderate bands. The *AQIPredict* LR method performs the best for RMSE (Fig. 5a) and band accuracy (Fig. 5c), while the *Pollutant2AQI* SVR performs the best for MAPE (Fig. 5b). Overall, the *Pollutant2AQI* methods have higher RMSE values but lower MAPEs. This shows that *Pollutant2AQI* methods can perform accurate predictions when the actual values are small; however, for points where actual values are large, the predicted values of *Pollutant2AQI* methods are further from the actual values than those of other methods, which results in large RMSE values but still small MAPE values.

Poles Lane and Ntl Physical Lab are two similar sites, which monitor the same three meteorological features and two kinds of pollution data: $NO_2$ and $O_3$ (dominant). Boxplot figures in Figure 3 show that their AQIs' distributions are also similar. Compared to the other sites, RMSEs of these two sites are larger, band accuracies are smaller, but MAPEs do not show much difference. An interesting finding is that *AQIPredict* NARX performs the best for the RMSE and MAPE evaluations for both sites, but *Pollutant2AQI* NARX has a better band accuracy than *AQIPredict* NARX. For Poles Lane, *Pollutant2AQI* NARX achieves the best band accuracy, while for Ntl Physical Lab, band accuracy is about 5% lower than those of Poles Lane, and *Pollutant2AQI* SVR achieves the best band accuracy.

The Marylebone Road kerbside site measures three meteorological features and five kinds of pollution data: $NO_2$ (dominant), $PM_{10}$, $O_3$, CO and $SO_2$. The majority of the AQI values of this site are close to 50, which is the boundary between the Good and Moderate band. However, the maximum AQI values reach the Hazardous band, i.e., the values cover the entire range of the 6 AQI bands; from Good to Hazardous. For the prediction performance for this site, *AQIPredict* NARX achieves the best RMSE, *Pollutant2AQI* NARX achieves the best MAPE, while *AQIPredict* LoR achieves the best band accuracy.

To summarise, for RMSE, *Pollutant2AQI* NARX and *AQIPredict* NARX perform the best on datasets from three sites each, with *AQIPredict* LR showing the best performance on the seventh case. For MAPE values (see Fig. 5b), *Pollutant2AQI* NARX performs the best on datasets from four sites, *AQIPredict* NARX performs the best on two, and *Pollutant2AQI* SVR performs the best on one. It is a mixed picture for band accuracy as shown in Fig. 5c, with *Pollutant2AQI* NARX showing the best performance for three datasets, *AQIPredict* LR, *AQIPredict* LoR, and *Pollutant2AQI* SVR separately showing the best performance on one dataset each, and *Pollutant2AQI* LR and *Pollutant2AQI* LoR tied in for similar accuracies on the last one. Taking into account all the datasets from the seven sites, *Pollutant2AQI* NARX performs the best on most of the datasets, and provides competitive results for the rest. This indicates that *Pollutant2AQI* NARX has robust performance for different kinds of datasets and can be recommended for AQI prediction.

### B. POLLUTANT PREDICTION: RESULT AND DISCUSSION

In addition to AQI prediction, we also compared MAPEs for the prediction of the individual pollutant values (as part of the *Pollutant2AQI* approach) by the different methods, i.e., LR, LoR, SVR, and NARX. The results are presented in Figure 6. We get the worst performance with LoR as the training algorithm across most of the datasets, with the only exception being the MAPE results for $PM_{10}$ data from Horley and the CO data from Marylebone Road (second lowest MAPE value). For $NO_2$, the proposed NARX approach performs the best for 6 sites, while SVR performs the best on data from Belvedere West. Both SVR and NARX get the same MAPE on $NO_2$ data from Marylebone Road. However, the NARX method does not appear to be the best one for predicting $PM_{10}$ data. Among the four sites monitoring $PM_{10}$ concentrations, LR achieves the two best MAPEs, while LoR and SVR achieving the best MAPE values on one dataset each. For $O_3$ data, NARX performs the best for two datasets, with LR and SVR performing well on one each. SVR also performs the best on one $PM_{2.5}$ dataset with NARX performs the best on the other one. NARX performs well for both $SO_2$ and CO datasets.

Overall, NARX can achieve a good performance for prediction of pollution data except for that of $PM_{10}$. Therefore, for predicting AQIs, NARX can be used on areas whose dominant pollutant is not $PM_{10}$, with LR proving to be a better choice for such locations. This is in agreement with findings in the existing literature [16], where multiple linear regression models achieved better results than ANN for mean relative and absolute error percentages as well as for RMSE for $PM_{10}$ concentration predictions.

### VI. CONCLUSIONS AND NEXT STEPS

In this paper we propose two approaches for AQI estimation and prediction, both based on meteorological and historical pollutant data; one learns a model based on the previous AQI and meteorological data to predict AQIs, the other learns models based on the previous pollution data and meteorological data to predict pollution data first and then compute AQIs. Both approaches can get good band accuracy (over 75%), as shown on the evaluations conducted across various datasets. The best approach is the latter approach combined with neural network, which achieves the lowest RMSE and MAPE across most of the evaluated datasets. This approach gets very good band accuracies (more than 81%) on all the datasets. However, by further analysing the individual pollutant value prediction step, we found that a neural network-based method is not the optimum at predicting $PM_{10}$ data. Therefore, we recommend using linear regression to predict AQI if the dominant pollution is $PM_{10}$ in the area of interest. In summary, the results show the feasibility of our proposed approaches for predicting AQIs based on meteorological data and the historical pollutant data/AQIs.

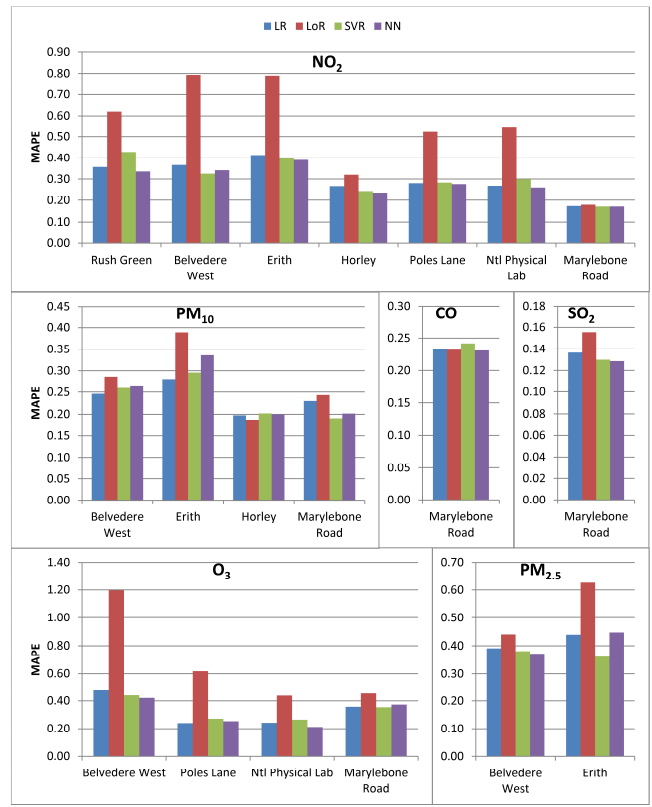

**FIGURE 6.** Results of Pollution Data Prediction of Different Learning Algorithms.

In the future, we plan to analyse correlations between sensing sites located close to each other to uncover latent similarities in pollutant or AQI patterns and to analyse if they are influenced by other environment factors such as green cover or traffic. We also plan to further extend the analysis of impact on air quality from different types of sensing areas across different cities. Another future work is to infer the latent diurnal and seasonal pollution data patterns in different parts of a city according to its built environment.

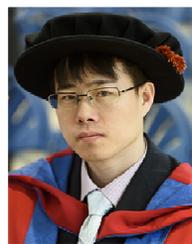

**YUCHAO ZHOU** received his Ph.D. degree in electronic engineering at the University of Surrey in 2018. He did his B.S. degree in telecommunications engineering with management from a joint programme between Beijing University of Posts and Telecommunications, China and Queen Mary University of London, UK, in 2011 and M.Sc. degree in communications networks & software from the University of Surrey, Guildford, UK, in 2012. He is currently a Research Fellow in the Institute for Communication Systems, at the University of Surrey. His research interests include semantic Web, search techniques for the Web of Things, and IoT applications in smart cities.

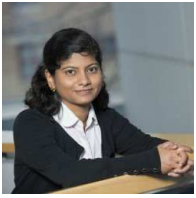 **SUPARNA DE** received her Ph.D. and MSc degrees in Electronic Engineering from the University of Surrey in 2009 and 2005, respectively. She is currently a Senior Research Fellow in the Institute for Communication Systems, at the University of Surrey. She has been leading technical work areas related to various aspects of service provisioning and data analysis in the Internet of Things domain in several EU projects such as TagItSmart, iKaaS, IoT.est, iCore and IoT-A. Her research has been supported by grants from the EC H2020 and FP7 programs and through DTI, UK-funded programs. Her current research interests include knowledge engineering methods, machine learning for data analytics, Web of Things and semantic association analysis. She is a member of IEEE and ACM.

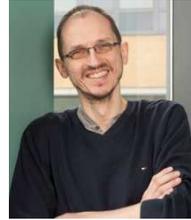 **KLAUS MOESSNER** is a Professor in the Institute for Communication Systems, at the University of Surrey. He was the founding chair of the IEEE DYSPAN Working Group (WG6) on Sensing Interfaces for future and cognitive communication systems. He was involved in the definition and evaluation of cooperation management between autonomous entities, in the UniverSelf project, and was technical manager of the iCore project and has the same role in the H2020 project CPaaS.io; he was project leader of IoT.est, SocIoTal, and currently leads the iKaaS project as well as working area 6, on System Architecture in the 5G Innovation Centre at the University of Surrey. His research interests include cognitive networks, knowledge generation, as well as reconfiguration and resource management and he is a senior member of IEEE.

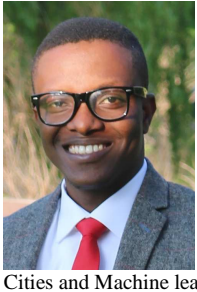 **GIDEON EWA** obtained his B.Eng degree in Computer Engineering at University of Uyo, Nigeria and his M.sc degree in Mobile and satellite communication from the university of Surrey, United Kingdom in 2016. He is currently a Satellite System Engineer with Center for Satellite Technology Development (CSTD), a Research Center under National Space Research and Development Agency (NASRDA) in Abuja, Nigeria. His research interest is in Wireless communications, Internet of things (IoT), Smart Cities and Machine learning.

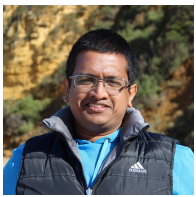 **CHARITH PERERA** (M'14) is a Lecturer (Assistant Professor) at Cardiff University, UK. He received his BSc (Hons) in Computer Science from Staffordshire University, UK and MBA in Business Administration from the University of Wales, Cardiff, UK and Ph.D. in Computer Science at The Australian National University, Canberra, Australia. Previously, he worked at the Information Engineering Laboratory, ICT Centre, CSIRO. His research interests are Internet of Things, Sensing as a Service, Privacy, Middleware Platforms, and Sensing Infrastructure. He is a member of both IEEE and ACM. Contact him at www.charithperera.net or charith.perera@ieee.org.